Authors version – accepted for publication in the Proceedings of the 2017 CIVEMSA conference – Annecy, June 2017# Quality Prediction in Injection Molding

Neural networks geometric dimension prediction on raw signals and thermographic images

Pierre Nagorny, Maurice Pillet, Eric Pairel
SYMME Laboratory
Savoie Mont Blanc University
Annecy, France
pierre.nagorny@univ-smb.fr,
maurice.pillet@univ-smb.fr,
eric.pairel@univ-smb.fr

Ronan Le Goff, Jérome Loureaux, Marlène Wali
IPC - Centre Technique Industriel de la Plasturgie et des Composites
Bellignat, France
ronan.legoff@ct-ipc.com

Patrice Kiener
InModelia
Paris, France
patrice.kiener@inmodelia.com*Abstract*— Injection molded part quality can be improved by precise process adjustment, which could rely on in-situ measurements of part quality. Geometrical and appearance quality (visually and sensory) requirements are increasing. However, direct measurement is often not feasible industrially. Therefore, process control must rely on a prediction of parts quality attributes. This study compares prediction performances of diverse neural networks architectures with "classical" regression algorithms. Dataset comes from inline industrial measurements. Regression was performed on 97 scalar statistical features extracted from multiple acquisitions sources: thermographic images and analog signals. Haralick features were extracted. Convolutional Neural Networks were trained on thermographic images and Long Short Term Memory networks were trained on raw signals. Although the dataset was small, neural networks show better predictions scores than other regression algorithms.

*Keywords— injection molding; quality prediction; thermography; Convolutional Neural Networks ; LSTM*## I. Introduction

Thermoplastics injection molding allows the production of complex parts in industry. Multiple parameters should be regulated and adjusted to achieve optimal final part quality. Final quality depends on multiple successively applied physical factors, from plastic melt temperature to injection process timing. Literature shows diverse methods for quality control which used adaptive learning algorithms and recurrent neural networks [1] . Final objective is to adjust the process from a characterized produced part to the next one, in about a short thirty seconds' cycle time. This must include quality measurement, adjustment computation and command set. Direct measurement of the part is often not achievable in an industrial context. Thus, part quality must be predicted from indirect achievable measurements.

Weight and dimension as quality attributes are recurrently predicted in literature [2]. An extensive research field is open by recent neural networks developments and successes. Injection molding adaptive control require prediction of multiple quality characteristics. In this study, we compare two methods for quality prediction: regression algorithms on extracted descriptors and neural networks on raw images and signals (Fig. 1) to achieve prediction of a continuous geometrical attribute of a produced part.

## II. Industrialy produced parts measurements

### A. Experimental data acquisition on inline industrial process

Plastic injection process has more than thirty factors, from material property to external humidity, all of which are time dependent. An experimental trial was held at the IPC Center (Bellignat, France) in industrial conditions. We produce 204 100x100 millimeters box with different machine adjustments. The injection process has a small capability window to achieve completion of the injected part. Thus, we choose adjustments to maximize variance across the process capability window. A pressure sensor and a temperature sensor were placed inside the mold. Additionally, we acquired on the machine the hydraulic injection pressure and the screw position. Production cycles of 30 seconds were records for each signal at 100 Hertz (Fig. 2).

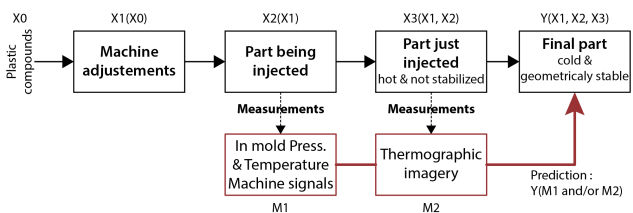

Fig. 1. Prediction of a final quality attribute with thermography, in-mold signals and machine signals

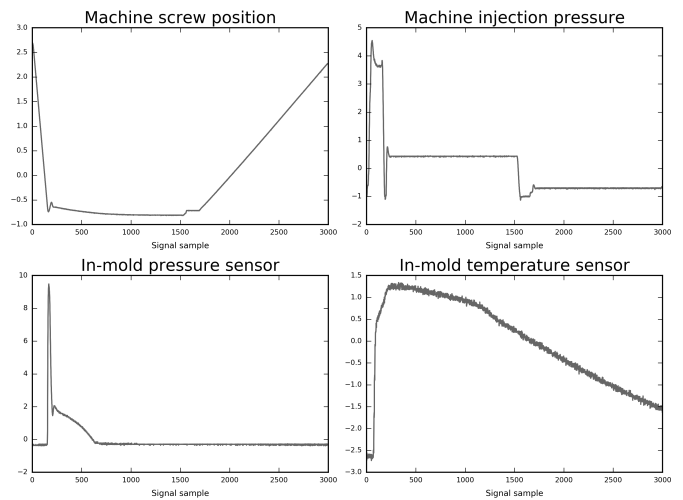

Fig. 2. Standardized process signals

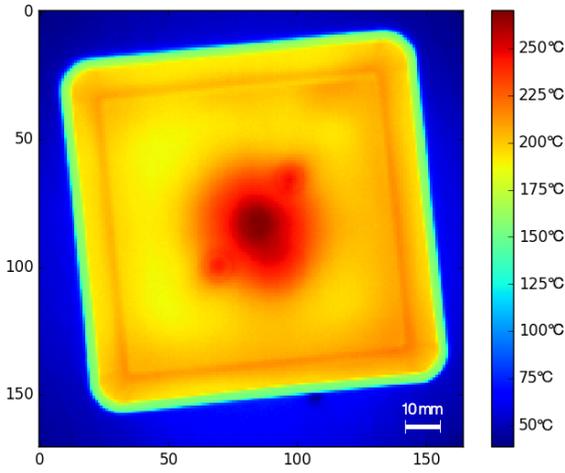

Fig. 3. Thermographic image of the injected part 10 seconds after production

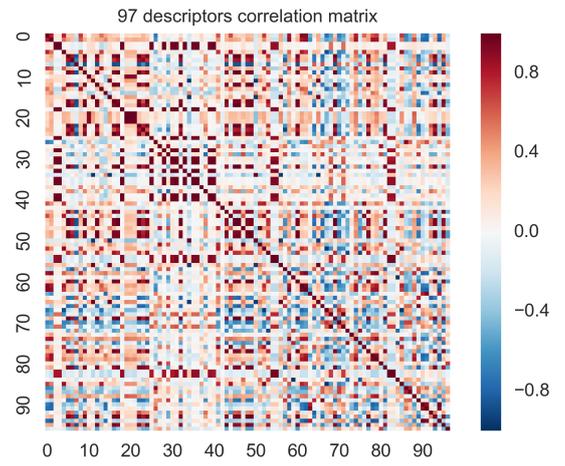

Fig. 4. Correlation matrix of 97 descriptors

A thermal camera was set-up in line. Though, we didn't use the continuous video measurements capability, a still frame was acquired for each part 25 seconds after the ejection of the part. We highlight on the ease of industrial setup of such measurements. Injection molding machines are equipped with sensors for process regulation. Some molds are already equipped with an internal sensor. Without sensor, a thermal camera is easier to set up on an industrial production, with a minimum cost. Discrete geometrical features measurements were performed ten days after production with a micrometer comparator. Quality characteristics responses were selected in term of industrial quality requirement. A part dimension is here studied as a predictable response: the width of the box. This dimension varies during the cooling and stabilization of the part and depends on the machine adjustments.

### B. Statistical descriptors extraction

Signals peaks were extracted using Continuous Wavelets Transform (CWT) [3] on raw signals (Fig. 2). Statistical descriptors were extracted from CWT, raw signals and thermographic images (Fig. 3):
- low order statistics: mean, median, standard-deviation, minimum, maximum, quantile 75 and 90;
- higher orders: mode, skewness, kurtosis.

Haralick parameters [4] were computed for each image. 26 thermographic statistical descriptors were extracted from 204 produced parts images. 71 statistical descriptors were also extracted from signals and CWT peaks.

### III. REGRESSIVE MODELS ON STATISTIAL DESCRIPTORS

#### A. Descriptors correlation analysis

Recursive features eliminations on linear regression $R^2$ scores shows that selecting 15 out of 26 thermographic descriptors produced the best score. The same algorithm on signals descriptors shows that selecting 30 features out of 71 produced the best $R^2$ score. Finally, selecting 46 descriptors out of the total 97 produced the overall best score. Correlation matrix of the 97 descriptors are shown in Fig. 4 (using the SeaBorn library [5]). Forty descriptors are correlated, mostly descriptors from raw signals and peak CWT transform.

#### B. Model comparaisons

We compare different regressive models using the Scikit-Learn library [6]. Hyper-parameters were optimized with cross validation. We compare Mean Square Error (MSE) and R squared ($R^2$) regression scores using extracted signals descriptors, thermographic images (*Thermo*) descriptors and both. Dataset was randomly split in 177 parts for training and 27 parts for testing from the 204 images. This dataset is small and this work can certainly be improved by using a larger set. Selected algorithms are : Least Absolute Shrinkage And Selection Operator (LASSO) [11], Elastic-Net [6], Support Vector Regression [8], Random Forest [9][10], K Nearest Neighbor [11], Stochastic Gradient Descent [12], Stochastic Gradient Boosting [13], Gradient Boosting regression [14], Decision Tree [15], Bagging Decision Tree [16], Ada Boosting Decision Tree (AdaBoost.R2 algorithm) [17]. Tuned hyper-parameters are shown in Table I. $R^2$ regression scores are shown in Fig. 5.

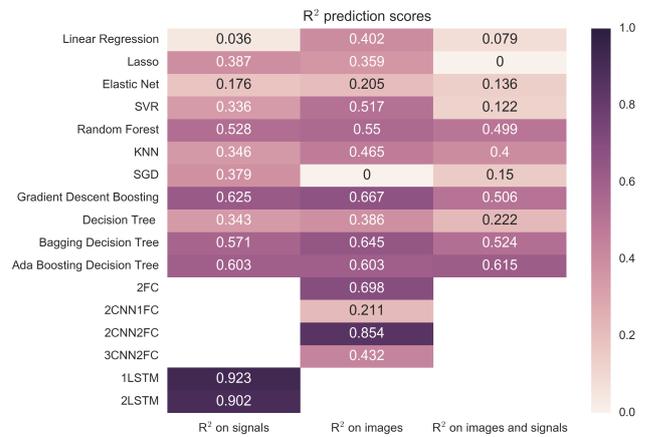

Fig. 5. Prediction score comparaison

TABLE I. MODELS TUNED HYPER-PARAMETERS

| Regressive model | Tuned hyper-parameters |
|---|---|
| Linear regressor | Only linear coefficients |
| Lasso | Regularization = 0.3162* L1 |
| Elastic Net | Regul = 0.00023*L1, 0.00033*L2 |
| Support Vector Regressor | Linear kernel, penalty parameter C = 1.0 |
| Random Forest | 100 trees |
| K Nearset Neighbor | Neighbor number = 2, KDtree algorithm, Minkowski metric with L1 distance |
| Stochastic Gradient Descent | Mean squared loss, L2 penality, regul = 0.0001*L1+0.00067*L2, 5 iterations, learning rate=0.01 / $t^{0.25}$ |
| Gradient Descent Boosting | 500 boosting stages with L1 loss, maximum depth = 4, samples to split = 2 |
| Decision Tree regressor | maximum depth = 1, MSE to measure split quality, samples to split=2 |
| Bagging Decision Tree | 10 Decision Tree estimators trained on 10 subsets, average final prediction |
| Ada Boosting Decision Tree | 300 Decision Tree estimators trained with linear loss with a 1.0 learning rate |
| Two Fully Connected Layers | regul = L2, L1 loss, learning rate = 0.001 Adam optimizer, 10 000 iterations |
| Two 5x5 CNN layers, 1 FC | |
| Two 5x5 CNN layers, 2 FC | |
| Three 3x3 CNN and 2 FC | |
| One LSTM layers | regul = L2, L1 loss, learning rate = 0.001 Adam optimizer, 100 000 iterations |
| Two LSTM layers | |

### C. Regression scores analysis

Some dataset features are highly correlated. Non-linear and complex interactions can be modeled by non-linear or iterative regressions. Best results are obtained for images descriptors and Gradient Descent Boosting regressor. Thermographic imaging of hot parts just after production contains more information for geometrical prediction than in in-mold sensors signals. However, using signals and images descriptors for regression shows bad results as features are too correlated.

TABLE II. MODELS TUNED HYPER-PARAMETERS

Prior features selections must be performed. Raw signals are functional of complex energetic thermal interactions of melt plastic and conductive metallic mold. Thus Earth Moving Distance [18] comparison could show good descriptive results, as an energetic descriptor. Works on thermal imagery features extraction have been done in breast thermographic imagery [19][20] and in material subsurface defect detection [21]. Thermal images descriptors could be completed with texture models tuned hyper-parameters features based on variance and entropy or SIFT and SURF. Descriptors could also be extracted from Fourier Transform and histograms comparisons. Furthers works can profit of the neural networks research field.

## IV. CONVOLUTIONAL NEURAL NETWORKS ON THERMOGRAPHIC IMAGES

### A. Raw datasets preprocessing

Contrary to the previous step where we extracted descriptors from signals and images, neural networks infer these descriptors with iterative backpropagation learning. Dataset used the same validation split: 177 parts for training and 27 parts for testing. This is a small dataset for training neural networks. Raw images of 156x156 pixels were reduced to 28x28 for training and standardized. This helps gradient convergence on our size limited training set. Furthermore, we could have use whitening or principal component analysis to decorrelate images. Images are single pixel valued, thus equivalent to grayscale: input tensor has a 28x28x1 shape.

### B. Multi Layers Perceptron networks architectures

Previous works [22] used a three layers MLP architectures (9-21-2) trained on 162 training parts. This achieved a 0.80 $R^2$ score on parts weight prediction and a 0.882 $R^2$ score on parts dimension prediction. All of our work was done using the TensorFlow library [23]. Firstly, a simple MLP [24] of 2 Fully Connected layers (2 FC) was trained. First layer transforms the 28x28x1 vector input into a 128x1 vector; then the second layer computes the final continuous value which predict the standardized width of the produced part. We obtained a 0.26 MSE and 0.70 $R^2$ on the test dataset. These are better scores than our previous regressions on descriptors and this is improvable.

### C. Convolutional Neural Networks architectures

A successful architecture for image processing are Convolutional Neural Networks [25] (CNN). We evaluated multiple architectures with one to three convolutional layers, one and two fully connected (architecture is shown in Fig. 6 and results are shown in Fig. 5). Hyper-parameters were tuned for each network. State-of-the-art networks are going deeper by stacking layers, but a larger dataset is then clearly needed for training. Two 5 by 5 pixels' convolution layer are successively applied to the 28x28x1 image input vector. Then 2 Fully Connected layers transform the 3136x1 vector into a 1024x1 vector after FC1 and to the final continuous predicted value after FC2. Network architecture is shown in Fig.3 using Tensorboard. Maxpooling (2x2 width and 2 stride) after each convolutional layers and Rectified Linear Unit (ReLu) [26] before the Fully Connected layer were evaluated and validated. Maxpooling reduced dimensions and speed up descent training, without losing much information; then ReLu as an activation function increased the non-linearity of the convolutional transforms, before the Fully Connected layers transforms to the continuous response: the predicted width of the part.

### D. Hyperparameters tuning

Batch size was chosen to fit with dataset. Best results were obtained with 17 images randomly chosen for each batch.
The common L1 distance for regression problem was used as a loss function. Mean Error, Mean Square Error, Root Mean Square and Huber loss were also evaluated but convergence was slower. To reduce overfitting, a 0.01 L2 penalty was set on backpropagation and a 0.90 dropout probability was set before the Fully Connected layers. Learning rate was set to 0.001 per iterations which is lower enough to guaranty gradient descent convergence. Stochastic gradient descent and RMSprop optimizer [27] algorithms were evaluated but best results were achieved with Adaptive Momentum stochastic gradient descent optimizer (Adam) [28] with suggested defaults parameters: 0.9 for first moment exponential decay, 0.999 for second moment and 1E-08 for epsilon stabilization term.

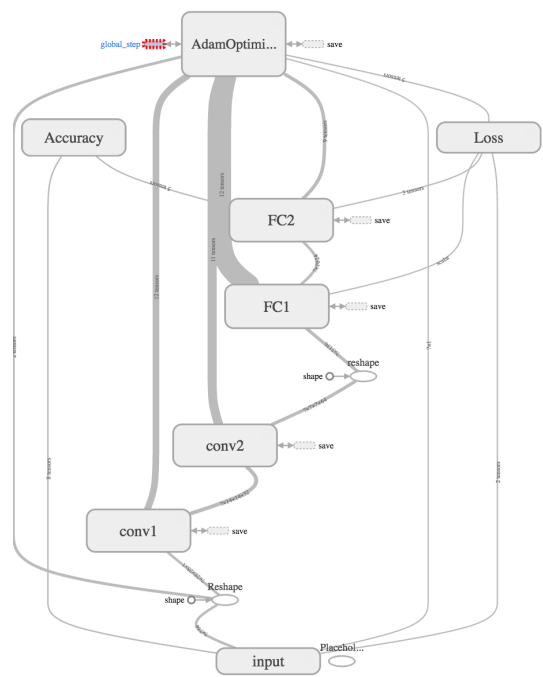

Fig. 6. 2 CNN 2 FC network architecture graph in TensorBoard

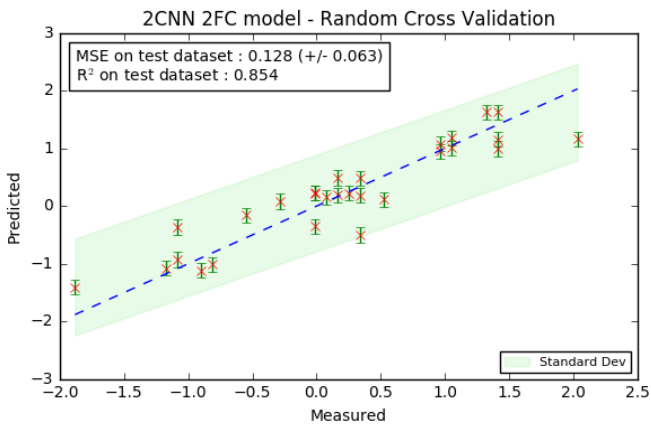

Fig. 7. Best convolutional model cross-validation

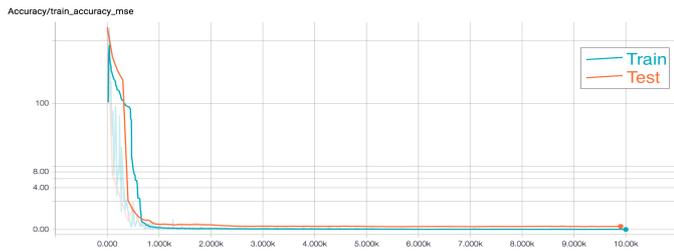

Fig. 8. 2CNN 2FC2 convergence : L1 loss and MSE

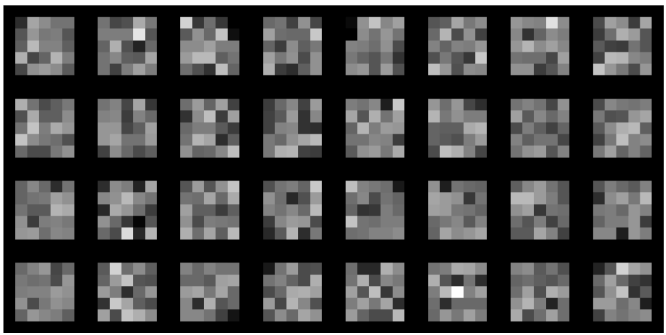

Fig. 9. 2CNN2FC model : trained convolutional layers (1st layer only)

Weights were initialized using normal randomized values. Xavier initialization [29] was tested but no performance improvement was observed. Convergence was observed after 1500 iterations (6 minutes); thus, early stopping was performed at 10 000 (40 minutes) to achieve the best generalization and to limit overfitting. Training iterations loss are shown in Fig. 8. A fast convergence is followed by a slight improvement.

*E. Results analysis*

In comparison with regression on multiple descriptors, we obtained a 0.13 Mean Squared Error on the test set and 0.85 $R^2$ correlation. Best results were obtained with a 2 CNN 2 FC networks (Fig. 7). Using shorthand notation, best architecture is C(32, 5, 1)-P-C(64, 5, 1)-P-D-N-FC(3136)-FC(1024), where C(d, f, s) indicates a convolutional layer with $d$ filters of spatial size $f \times f$, applied to the input with stride $s$. D is a dropout layer. N is a non-linear ReLu layer. FC(n) is a fully connected layer with $n$ nodes. The trained weights of convolutional filters were extracted for visualization (inspired by [30] the first layer is shown in Fig. 9). Filters started to react to real patterns, but generalization is not achieved because of the small training dataset. L2 regularization and ReLu did not limit overfitting on this small architecture. Data augmentation with a deeper architecture, and advanced preprocessing could certainly increase generalization. We will investigate networks retraining and Generative Adversarial Networks [31] on this small dataset problem; but the most encouraging research direction is multimodal networks fusion, as another signals dataset is here available.

V. LONG SHORT TERM MEMORY NETWORKS ON RAW SIGNALS

*A. Simple datasets preprocessing*

We worked with the raw analog signals, directly acquired from the machine or sensors. Signals slightly vary in length due to industrial recording. Dynamics networks were not used in this work; thus, signals were resampled to 3000 sample points. Dataset was then standardized for training purpose, but no filter was applied.

*B. Recurrent Neural Network architecture*

We use a recurrent network with Long-Short Term Memory (LSTM) cells [32] to train a predictive model of the part width. Fig. 10 shows the LSTM with one layer network architecture**.** Previous works showed good results on raw signals classification [33]. We use the TF.Learn library [34] to evaluate two architectures: with only one LSTM layer and with two LSTM layers. Mean Score Error loss and the previous successful Adam optimizer were used for training. A classical 0.001 learning rate was used with 100 000 training iterations.

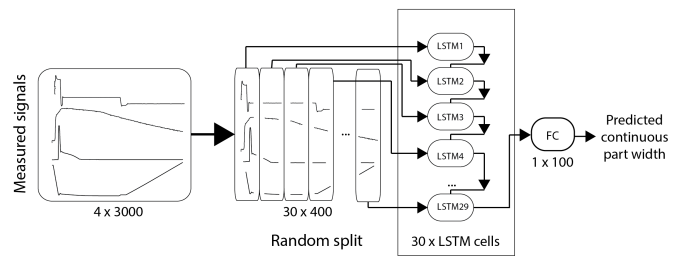

Fig. 10. One LSTM architecture

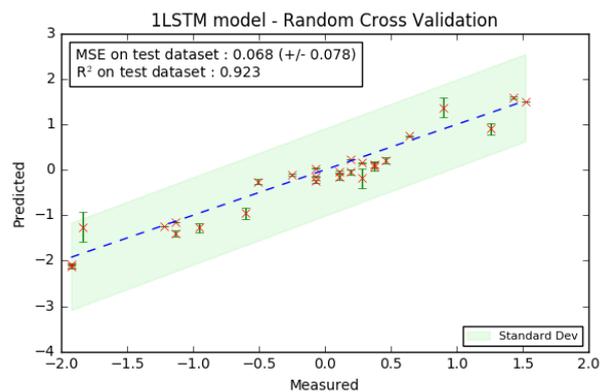

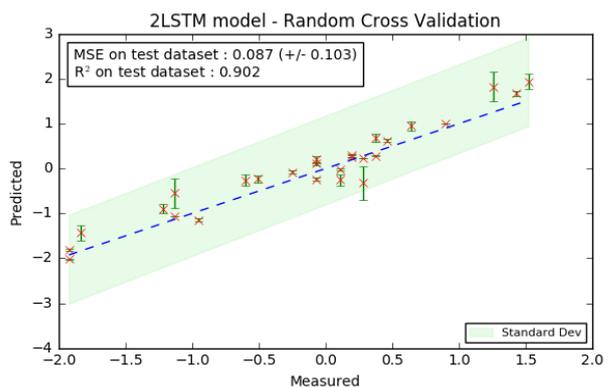

Fig. 11. 1 LSTM and 2 LSTM networks cross-validation scores

We observe the 30 cells which received a randomly splitted part of the training batch signals. The dual LSTM layers has the same architecture but with two parallel and splitted signals parts. This aims to learn different features and should help generalization. Furthers works should evaluate recent recurrent networks architecture and compare them with non-neural networks based time series machine learning algorithms.

*C. Prediction scores analysis*

Results (Fig. 11) show better performance for the unique LSTM layer architecture. This is probably due to the small training dataset, which increased overfitting on a small network. Recurrent networks architecture is also complex. Specific dataset preprocessing must be study. However, scores are higher than all previously evaluated prediction algorithms and prediction systems from the literature. Results are encouraging. Furthers works will be held on recurrent network tuning for this specific industrial application.

## VI. RESEARCH PERSPECTIVES

Our results show neural networks prediction potential with a 0.92 $R^2$ score on raw signals and 0.85 on images. Scores could be improved by networks tuning, using Parametric Rectified Linear Unit (PReLU) [35] for smoothed gradient descent, and optimized early stopping while training. The fusion of both images and signals best predicator will certainly achieve an even better score. Multimodal fusion has been achieved with audio and video data [36]. Fusion of a temporal and a spatial video stream convolution pipeline showed nice results, by using a simple average late fusion of the prediction layer outputs [37]. Neural networks seem to show better generic performance on complex and correlated nonlinear problems. Injection process. More complex networks architectures could be optimized but this is time-consuming. However, setting up a neural network architecture takes probably the same time as developing application specific images and signals processing system. The most important work for neural network training is to build a large dataset. The size of the dataset is the main limit of our present study. An open source and collaborative dataset for solid mechanics engineering would help to develop and validate new neural networks architectures. Adaptive control could rely on a prediction of the produced part quality. In this paper, we compare algorithms to predict a unique continuous characteristic. Furthers works will study multiple quality characteristics prediction. We study geometrical quality which is crucial for technical parts but other sensorial qualities (haptic and visual) are also decisive industrial challenges. Quality measurements can, and should, come from diverse technologies. An adaptive control system must then compute multiple machine adjustments based on multiple predicted characteristics of the next produced part. Neural networks show potential for multimodal data fusion and predictive control as neural networks show better predictions scores than other regression algorithms on a small dataset.